\documentclass[aps,floatfix,showpacs,superscriptaddress,twocolumn]{revtex4} 

\usepackage{color,amsmath}

\definecolor{purple}{rgb}{0.5,0,0.5}

\usepackage[colorlinks=true, pdfstartview=FitV, linkcolor=purple, 
citecolor= purple, urlcolor=blue]{hyperref}

\def\fs{\; \; .}

\newcommand{\lsp}{\hspace{0.03em}}

\newcommand{\QCD}{\sc{q\lsp c\lsp d}\rm}
\newcommand{\chpt}{$\chi$\lsp{\sc p\lsp t}\rm}
\newcommand{\<}{\langle}
\renewcommand{\>}{\rangle}

\newcommand{\bce}{{\sc b\lsp c\lsp e}\rm}
\newcommand{\FS}{{\sc f\lsp s}\rm}
\newcommand{\cO}{\mathcal{O}}

\begin{document}

\begin{flushleft}
IFIC/07-18\\
UWThPh-2007-12
\end{flushleft}

\title{On the mesonic Lagrangian of order $\mathbf{p^6}$ in chiral 
$\mathbf{SU(2)}$}

\author{Christoph Haefeli}
\affiliation{Departament de F\'{\i}sica Te\`orica,
         IFIC, Universitat de Val\`encia -- CSIC,
         Apt. Correus 22085, E--46071 Val\`encia, Spain}
\author{Mikhail A. Ivanov}
\affiliation{Bogoliubov Laboratory of Theoretical Physics, 
Joint Institute for Nuclear Research, 141980 Dubna, Russia}
\author{Martin Schmid}
\affiliation{Institute for Theoretical Physics, University of Bern,
Sidlerstr. 5, CH-3012 Bern, Switzerland}
\author{Gerhard Ecker}
\affiliation{Faculty of Physics, University of Vienna, Boltzmanngasse
  5, A-1090 Wien, Austria}

\begin{abstract}
We show that the number of operators in the presently known mesonic chiral
Lagrangian of order $p^6$ in the two--flavour sector can be reduced by at
least one from 57 to 56 by providing an explicit relation among the
operators. We briefly discuss the relevance of this new relation.
\end{abstract}

\pacs{11.30.Rd, 11.40.Ex, 12.39.Fe}
\maketitle


\noindent
Chiral perturbation theory (\chpt) \cite{weinberg,glann,glnpb} is the
effective field theory of the strong interaction, \QCD, at low
energies. It relies on an effective
Lagrangian that parametrises the most general local solution of the chiral Ward
identities in a long--distance expansion. In the mesonic sector, the
Lagrangian up to $\cO(p^4)$ had been constructed by Gasser
and Leutwyler in \cite{glann,glnpb}.

The advent of
next--to--next--to--leading--order calculations in \chpt{} called for the
construction of the effective chiral Lagrangian in the mesonic sector at
$\cO(p^6)$. This investigation was first performed by Fearing and
Scherer (\FS{}) \cite{Fearing:1994ga} and was later revisited by Bijnens,
Colangelo and Ecker (\bce{})\cite{Bijnens:1999sh}. We refer to \bce{} for a
detailed comparison with \FS. In short, the Lagrangian in
\cite{Bijnens:1999sh} consists of fewer chiral operators. Since it was shown
there that one may express all monomials of \FS{} in the basis of \bce, this
implied that the Lagrangian of \FS{} can not be minimal.

To make the main point of this comment clearer later, we 
repeat the procedures that guided \cite{Bijnens:1999sh} during the
construction.  In a first step one writes down all possible
(products of) traces of products of chiral invariant operators with
total chiral dimension six that are Lorentz--, $C$-- and $P$--invariant as well
as hermitian. It is not necessary to restrict the number of light flavours at
this stage, however. In a second step the following relations or
procedures may be used to obtain a minimal set of independent monomials at
$\cO(p^6)$ for $n$ flavours:

\begin{itemize}
\item[i.]   Partial integration in the chiral action of $\cO(p^6)$;
\item[ii.]  Equation of motion;
\item[iii.] Bianchi identity.
\end{itemize}
These simplifications and the justification of their use are
described in detail in \bce{} and we refer to this article for more
details. If one restricts oneself then to a particular case of $n$ -- of
relevance are $n=2$ and $n=3$ -- one notices that for each $n$ there
are additional linear relations among the invariants of $\cO(p^6)$ due to the
Cayley--Hamilton theorem. E.g., for $n=2$ it implies
\begin{equation}
  \label{eq:CaHa2}
  \{A,B\} = A\,\<B\> + B\<A\> + \<AB\> - \<A\>\<B\>
\end{equation}
for arbitrary two--dimensional matrices $A,B$; with $\<\cdot\>$ we denote the
trace of the enclosed matrix. Despite the careful analysis in \bce{} (who took
  Cayley--Hamilton relations into account), we realised that there
exists at least one additional relation for $n=2$ among their basis
elements $P_i$. Adopting the notation from \bce{}, we find via the
procedures i.--iii. as well as the Cayley--Hamilton theorem
\begin{equation}
\begin{split}
  \label{eq:nonminimal}
  8&P_{1}
      -2P_{2}
      +6P_{3}
      -12P_{13}
      +8P_{14}
      -3P_{15}
      -2P_{16}\\
      -20&P_{24}
      +8P_{25}
      +12P_{26}
      -12P_{27}
      -28P_{28}
      +8P_{36}
      -8P_{37}\\
      -8&P_{39}
      +2P_{40}
      +8P_{41}
      -8P_{42}
      -6P_{43}      
      +4P_{48}=0
\fs
\end{split}
\end{equation}
Several comments are in order:
\begin{itemize}
\item[-] Obviously, Eq.~(\ref{eq:nonminimal}) allows to drop one
  of the $\mathrm{SU}(2)$ basis elements of \bce{}. 
  A natural choice would be $P_{27}$ whose low--energy constant, to our
  knowledge, has not occurred in any of the  
  matrix elements calculated so far in the literature. 
\item[-] We emphasise that the derivation of Eq.~(\ref{eq:nonminimal})
  required no additional algebraic manipulations that were not already
  used by \bce{}. As the example shows,  it
  is a nontrivial matter to ensure minimality of a
  given Lagrangian in \chpt{} at higher orders.
\item[-] One would like to know if the Lagrangian at
  order $p^6$ with two flavours is by now minimal. The answer requires 
  a thorough investigation and we postpone it to a later
  publication. 
\item[-] Since in \bce{} the Lagrangians for $n=2$ and $n=3$
  were constructed in
  close relation, one might also question the minimality of the
  latter one. However, we did not find any indication of additional relations
  among the basis elements there. 
\item[-] The authors of \bce{} provided also the complete
  renormalization of \chpt{} at order $p^6$ \cite{Bijnens:1999hw}. It is
  straightforward to implement the additional linear relation into the
  renormalization prescription given there.
\end{itemize}


\section*{Acknowledgements}

\noindent

We thank Gilberto Colangelo for providing us with notes that were useful to
derive the new relation and
J\"urg Gasser for discussions, continuous support and comments on the
manuscript. 

\noindent
The calculations were performed using Form 3.1 \cite{Vermaseren:2000nd}.

\noindent
This work was supported by the Swiss National
Science Foundation, by Ministerio de Educaci\'on y Ciencia under the
project \sc fpa\rm \oldstylenums{2004}-\oldstylenums{00996}, and {\sc
  eu\rm} contract \sc mrtn-ct-\rm\oldstylenums{2006}-\oldstylenums{035482}
(\sc flavia\it net\rm). 


\end{document}